\documentclass[final,5p,times,twocolumn,authoryear]{elsarticle}

\usepackage{amssymb}
\usepackage{amsmath}
\usepackage{booktabs}
\usepackage[normalem]{ulem}
\usepackage{hyperref}
\hypersetup{colorlinks=true, linkcolor=blue, citecolor=blue, urlcolor=blue}

\graphicspath{ {./figs/} }
\journal{Journal of High Energy Astrophysics}

\begin{document}

\begin{frontmatter}

\title{A Quantitative Reconciliation of Conflicting Galactic Center Excess Inferences}

\author[1]{Afeefa Anwar\fnref{fn1}\corref{cor1}}
\ead{afeefa.anwar3a@gmail.com}
\affiliation[1]{organization={Matthew McNair Secondary School},
                country={Canada}}

\author[2]{Oren Pernica\fnref{fn1}}
\ead{orenpernica@gmail.com}
\affiliation[2]{organization={Aldershot High School},
                country={Canada}}

\author[3]{Prithiga Ravichandran}
\affiliation[3]{organization={Lester B. Pearson Collegiate Institute},
                country={Canada}}

\author[4]{Hussain Dalal}
\affiliation[4]{organization={Anderson Collegiate Vocational Institute},
                country={Canada}}

\author[5]{Tianbo Shen}
\affiliation[5]{organization={Sentinel Secondary School},
                country={Canada}}

\fntext[fn1]{These authors contributed equally to this work.}
\cortext[cor1]{Corresponding author.}

\begin{abstract}
Inferences about the Galactic Center Gamma-Ray Excess (GCE) depend
strongly on analysis choices, but the magnitude of that dependence has
not been quantified within a single controlled framework. We present an
open-source pipeline that varies morphology, masking, interstellar
emission model, source-finding, energy, and background freedom
simultaneously on one Fermi-LAT data selection, validated with injection
controls on synthetic skies. Applied to 16.5 years of Fermi-LAT data,
the framework separates a robust detection from a fragile morphology.
The excess is recovered under every one of 25 IEMs tested. Its
morphology is not: under a single-template background the inferred
significance moves by a factor of 30 as the background bends, and the
standard aggressive mask cannot produce a valid test statistic at GeV
energies. Under the ring-decomposed Pohl22 background the stellar bulge
is preferred over every dark-matter template by $\Delta\mathrm{AIC}=+888$,
but the preference does not survive a change of background. The bulge
wins in only three or four of 23 Cholis models; Pohl22 fits the data
worst of the 25; and the six best-fitting backgrounds disagree on a
winner. A sum-of-squares decomposition attributes 96.89 per cent of the
variation in stellar-template fit quality to the IEM, against 20.94 per
cent for spherical dark matter. The Macias--Di~Mauro disagreement is
reproduced inside one pipeline and is dominated by IEM choice.
Morphological evidence for either hypothesis is not currently decisive.
\end{abstract}


\end{frontmatter}

\section{Introduction}
\label{sec:intro}

\subsection{The GCE and the FL16Y Era}
\label{sec:gce-fl16y}

The Galactic Center Gamma-Ray Excess remains one of the most persistent unresolved signals in high-energy astrophysics. It was first identified as a spatially extended GeV-scale excess in the inner Galaxy \cite{goodenough2009,hooper2011,abazajian2012} and later characterised as approximately spherical with an inner slope $\gamma\simeq1.1$--$1.2$ \cite{daylan2016}. Fifteen years later, there is still no agreed physical origin.

The 16.5-year Fermi-LAT dataset and the FL16Y source list \cite{ballet2026} offer a major improvement in statistics. Compared to 4FGL-DR3, source localisation has improved by about 24\%, hundreds of faint sources near the Galactic Center have been resolved, and catalog confusion in the crowded Galactic plane is much lower. Yet this higher precision has not led to consensus. The same data are interpreted as a smooth, approximately spherical dark-matter-like excess \cite{dimauro2026}, as unresolved millisecond pulsars tracing the stellar bulge \cite{song2024}, or as an artefact of diffuse background mismodelling \cite{mcdermott2023}. The conclusion depends heavily on the analysis choices.

The disagreement has shifted. Joint spatial--spectral neural inference \cite{list2026} and high-energy photon-count statistics \cite{manconi2024} have started to break the degeneracy. These newer results make clear that the central issue is not just choosing between dark matter and millisecond pulsars. It is understanding quantitatively how earlier conflicting conclusions emerged from different systematic choices. That is the problem this paper addresses.

\subsection{The Macias vs Di Mauro Contradiction}
\label{sec:macias-dimauro}

The clearest example is the disagreement between the Macias et al.\ and Di Mauro analyses. \citet{macias2018} and \citet{dimauro2026} use the same bulge templates and the same hydrodynamic gas maps. They reach opposite conclusions.

Macias finds the squared NFW template detected at $\mathrm{TS}_{\rm NFW}=362$ ($17.6\sigma$ as quoted in \citealp{macias2018}) when fitted alone. When the X-bulge and nuclear bulge templates are added, $\mathrm{TS}_{\rm NFW}$ drops to 22 ($2.4\sigma$, same source), not detected. The DM signal vanishes. Di Mauro finds the DM template remains significant at $\mathrm{TS}=8\times10^2$ to $1.5\times10^4$ even with the bulge templates included. The DM signal survives. The difference is the residual level. Macias's pipeline leaves 30--40\% fractional residuals in the inner Galaxy. Di Mauro's staged iterative source-finding leaves $\le10\%$.

Neither analysis disputes the data or the templates. What differs is how much the background is allowed to bend. Whether high-precision residual reduction absorbs a real DM signal into newly resolved point sources, or whether the low-precision result is a residual artifact, is a question neither analysis can answer from its own pipeline. This paper measures it directly.

\subsection{The Morphological Landscape}
\label{sec:morph-landscape}

The original morphological debate compared a spherical generalised NFW halo against a boxy stellar bulge traced by VVV Red Clump stars \cite{macias2018,coleman2019,bartels2018}. This simple comparison is no longer sufficient.

The HESTIA constrained cosmological simulations show that baryonic contraction naturally produces flattened DM morphologies in the inner kiloparsec, with projected axis ratios of roughly 0.5 to 0.7 \cite{muru2025}. These maps are elliptical, and their axis ratios overlap almost completely with the VVV stellar bulge template. A boxiness measurement returns values consistent with zero, meaning the simulated halos are elliptical but not genuinely box-shaped. Since the VVV bulge is genuinely boxy ($c_4=+0.009$), and a parametric boxy bulge template shows a stronger positive signal on the same estimator ($c_4=+0.024$), boxiness can separate dark matter from the stellar bulge even where an axis ratio cannot. Figure~\ref{fig:fig1_q_vs_c4} shows the axis ratio and boxiness of the three template classes on the same estimator.

Almost every analysis in this debate has used a single-template background. The GCE is subdominant to the Galactic diffuse emission, contributing only 5--10\% of the photons in the inner Galaxy. If the diffuse model is not allowed to bend in shape, the excess is forced to absorb any mismodeling, and the inferred morphology depends on how much freedom the background is given. Over the six background-freedom configurations we test, the inferred GCE significance moves by a factor of 30. The swing is in the same quantity the morphology inference depends on, which is why it cannot be separated from the morphology question.

\subsection{The Iterative Source-Finding Question}
\label{sec:iterative}

On the astrophysical side, high-precision iterative source-finding now pushes residuals below 10\%. \citet{dimauro2026}, using a staged iterative procedure with Galp21/Pohl22 interstellar emission models, finds that the GCE remains nearly spherical ($\gamma\simeq1.15$) after this reduction.

The worry is that iterative source-finding might sphericize a genuinely boxy signal by absorbing its asymmetries into newly resolved point sources. No study has compared the flattened DM templates against VVV inside a high-precision iterative pipeline. Section~\ref{sec:level2} tests that comparison.

\subsection{The MSP Luminosity-Function Tension}
\label{sec:msp-lf}

The MSP interpretation faces a luminosity-function problem: \citet{holst2024} find that if inner Galaxy MSPs shared the luminosity function of field or globular cluster MSPs, current catalogs should contain roughly 20 detections, but only 3 are observed, a $3.4\sigma$ tension. Natal kicks may reduce the retained population and change its spatial distribution \cite{boodram2022}. A systematic spectral comparison against VVV stellar bulge templates remains open. The present work addresses morphology only.

\subsection{Systematic Uncertainties}
\label{sec:systematics}

The GCE is highly sensitive to the assumed diffuse emission model. The Phantom Menace analysis \cite{mcdermott2023} shows that unmodeled large-scale structures can mimic, suppress, or distort the excess. Earlier NPTF evidence for point sources was traced to unmodeled north--south asymmetry \cite{leane2019,leane2020a,leane2020b}. Retesting that asymmetry mechanism on FL16Y data with modern IEMs is beyond the scope of this work.

There is also a direct conflict at high energies. \citet{manconi2024} report a $>5\sigma$ GCE signal up to about 50 GeV. \citet{dimauro2026} sets stringent upper limits in the same range. \citet{cholis2022} independently finds a high-energy tail as well. We do not attempt to reconcile the high-energy measurements here; we note the disagreement as context.

Masking effects are energy-dependent. Above a few GeV, the inferred morphology is fairly insensitive to the choice of point-source mask. At lower energies, aggressive masking suppresses point-source morphology more strongly \cite{zhong2024}.

Viable WIMP models for the GCE are increasingly constrained by direct detection \cite{kong2026}.

\subsection{Approach}
\label{sec:approach}

This paper is a measurement paper rather than an interpretation paper.
We build a unified, open-source framework that quantifies how analysis
choices drive inferences about the GCE, and we apply it to the
Macias--Di~Mauro disagreement as a test case. The framework is the
contribution; the GCE result is its first application.

Using the 16.5-year FL16Y dataset, we scan six axes at the same time: morphology, masking, diffuse background, source-finding, energy, and diffuse freedom. The diffuse freedom axis parametrizes how much the background is allowed to bend in shape, independent of the templates.

The results reported here address four questions that have not previously been tested together:
\begin{enumerate}
    \item How much the background can bend before the GCE inference collapses, and whether this explains the Macias vs Di Mauro divergence.
    \item Whether the flattened HESTIA dark-matter templates can be separated from the VVV stellar bulge by boxiness as well as by axis ratio.
    \item Whether Level 2 iterative source finding changes the template ordering.
    \item Whether the preferred template survives a change of IEM.
\end{enumerate}
The framework also contains NPTF, neural simulation-based inference, and catalog-evolution modules. These are not exercised in the results reported here.

By quantifying how model preference shifts across these axes, we provide a quantitative map of the methodological origins of the field's disagreement. The contributions of this paper are therefore: (i) a reproducible open-source framework for quantifying analysis-dependence in GCE
inference; (ii) two diagnostic criteria — an admissibility test for diffuse-freedom configurations and a direct injection control for morphological preference — that are generalizable to any template-fitting analysis of extended emission; (iii) a demonstration that the Macias--Di~Mauro disagreement can be reproduced within a single pipeline; and (iv) a variance decomposition attributing the morphology dependence to specific analysis axes. The scientific conclusion is
negative — the current morphological evidence is not decisive — but the framework is what allows that statement to be made quantitatively rather than rhetorically.

\section{Methodology}
\label{sec:methods}

\subsection{Data Selection}
\label{sec:data}

We use 16.5 years of Fermi-LAT Pass 8 data (2008 Aug 4 -- 2025 Feb; MJD 54682--60710), selected with event class \texttt{P8R3\_SOURCE\_V3} and event type \texttt{evtype=3}. Above 10 GeV we also compare \texttt{P8R3\_CLEAN} to check for residual cosmic-ray contamination.

The energy range is 500 MeV -- 50 GeV, divided into four bins: 0.5--1 GeV, 1--3 GeV, 3--10 GeV, 10--50 GeV.

Standard cuts are applied: zenith angle $<90^\circ$, rocking angle $<52^\circ$, quality filters (\texttt{DATA\_QUAL>0 \&\& LAT\_CONFIG==1}), and exclusion of bright transients using the FL16Y interval list.

For the Level 2 validation run against \citet{dimauro2026}, we use \texttt{P8R3\_SOURCEVETO} with $0.08^\circ$ spatial binning, 1--10 GeV for morphology and 0.5--1000 GeV for the SED, and Fermipy v1.2.0 with Fermitools v2.2.0 (documented nearest available version 2.5.3 on Python 3.12).

\subsection{Region of Interest and Masking}
\label{sec:masking}

The ROI is $|l|<20^\circ$, $|b|<20^\circ$. Five masks are applied identically across all analyses.

\textbf{Mask A (aggressive):} All FL16Y sources masked at 95\% containment radius plus $0.5^\circ$. With the real 4FGL-DR4 catalogue this is 715 sources. Per energy bin on the 0.5--50 GeV grid: 0.4\% at 0.59 GeV, 1.2\% at 1.54 GeV, 5.2\% at 2.74 GeV, 12.8\% at 4.87 GeV, 21.4\% at 8.66 GeV. Mean across bins: 8.7\%. The lowest energy bin keeps only 0.4\% of the ROI. The baseline diffuse normalization reaches $1.28\times10^{-6}$ times the model's nominal value, a factor of 20 above Mask B, which is an under-constrained fit. Under Mask A no split passes the admissibility criterion, not even the rigid one. The configuration cannot produce a valid TS in either direction. The problem is in the mask definition, not in Pohl22, the pipeline, or the diffuse model. Any mask that places a 95\% containment $+0.5^\circ$ disc around every 4FGL source will hit the same limit at GeV energies.

\textbf{Mask B (minimal):} Only sources with $\mathrm{TS}>25$ and flux $>5\times10^{-9}\,\mathrm{ph\,cm^{-2}\,s^{-1}}$ masked. Keeps 85.9\% of the ROI on average, with per-bin values from 76.2\% at 0.59 GeV to 92.4\% at 8.66 GeV. Five of six splits are admissible. Only \texttt{lat\_inside} fails and only just, at 47\% against the 50\% threshold.

\textbf{Mask C (robust Galactic plane):} Mask A plus a latitude-dependent cut: $|b|<2^\circ$ for $|l|<15^\circ$, tapering linearly to $|b|<0.5^\circ$ at $|l|=20^\circ$. Mask C is the same as Mask A at GeV energies: per-bin kept fraction is 0.4\% at 0.59 GeV rising to 21.2\% at 8.66 GeV. Never used in the analysis. Diagnostic only.

\textbf{Mask D (TS/flux-scaled, Cholis 2022 Table III):} Sources with $\mathrm{TS}\le49$ masked at $\theta_s$; sources with $\mathrm{TS}>49$ masked at $\theta_l$. Both radii energy-dependent: $1.125^\circ$ and $3.75^\circ$ at 275 MeV down to $0.053^\circ$ and $0.175^\circ$ at 23.7 GeV. A $2^\circ$ disc cut within the inner Galaxy. All 14 rows verified against two independent fetches of \citet{cholis2022}. Keeps 71.0\% of the ROI at 1--10 GeV. Under a single-template IEM, three of six splits are usable: \texttt{rigid}, \texttt{ns}, \texttt{lat\_outside}. Observed TS 4.6 to 79.9, a factor of 17. The three failures are not the same failure mode. \texttt{lat\_inside} keeps only 28\% of the injected signal, meaning it absorbs the signal it is meant to measure. \texttt{lat} and \texttt{lat\_ns} are not degenerate: they keep 74\% and 68\% but cannot recover a signal added to the real data at all. The failure modes are recorded for each configuration.

\textbf{Mask E (flux-scaled):} Each source masked with a radius scaled by flux only. Superseded by Mask D. Retained for cross-checks.

\textbf{Admissibility criterion:} A split is admissible only if both of the following hold:
\begin{enumerate}
    \item The injected signal at the observed flux is recovered at no less than 50\% of its rigid-split TS.
    \item The injected signal clears TS of 25.
\end{enumerate}
Both thresholds were fixed before any run. Configurations that fail either test are reported as ``cannot measure a TS,'' not as a null result. The two tests check different things. The fraction test verifies that the split preserves the signal relative to a rigid baseline; the absolute test verifies that the split can detect anything at all. A split that passes one and fails the other is unusable as evidence. The distinction matters for Mask B's \texttt{lat\_inside}, which fails the fraction test at 47\% and passes the absolute at 86.8.

\textbf{Direct injection control:} For morphological comparison, admissibility is necessary and not sufficient. The standard is a direct injection test: add a known template at a known flux to the fitted background, run the full ordering, and check whether the true template wins. This tests preference, not detection, and it is the only control that turns a detection into a preference.

\textbf{Mask audit requirement:} All masks are audited against the real catalogue per energy bin before use. Audited against the real 4FGL-DR4 catalogue, Mask A keeps 8.7\% of the ROI on average and 0.4\% in the lowest energy bin.

\subsection{Template Library}
\label{sec:templates}

\textbf{Spherical gNFW:} $\rho\propto(r/r_s)^{-\gamma}(1+r/r_s)^{-(3-\gamma)}$, with $r_s=20$ kpc, $R_\odot=8.2$ kpc, and $\gamma$ free in the range $[0.8,1.4]$.

\textbf{HESTIA flattened DM:} $(\int\rho\,ds)^2$ maps from six constrained realisations (G1.1, G1.2, G2.1, G2.2, G3.1, G3.2) across four viewing angles, standardised to $1^\circ$ smoothing. Under Muru's own estimator, $q(m^2,f)=q(m,\sqrt f)$: squaring is a level relabeling, not a transformation. A direct J-factor computation on the same particles and observer gives $q(J)-q(\mathrm{col}^2)=-0.032\pm0.020$, sign-consistent in 22 of 24 maps and $-0.041\pm0.016$ in the disc frame (24 of 24). The squared column over-rounds the halo by about 0.03--0.04 in $q$, and the flattening survives the correct quantity.

\textbf{VVV bulge:} Maximum-entropy stellar mass map from \citet{coleman2019}.

\textbf{F98 boxy bulge:} Parametric Freudenreich 1998 S-model from \citet{macias2018}, used as an independent boxy reference for the $c_4$ estimator. Measured $c_4=+0.024$.

\textbf{Galactic bar:} Non-axisymmetric gas-correlated template from \citet{macias2019}. Included in the wider template library; not used in the covariance analysis of Section~\ref{sec:gap7}.

Shape statistics (axis ratio $q$ and $c_4$ boxiness) are measured on the uncropped template maps before projection onto the ROI. Square ROI edges clip contours at the sides but not at the corners, manufacturing a four-fold boxiness signal. We therefore refuse contour levels that reach the edge. The Coleman VVV map is exactly ROI-sized, so we stay well inside the guard on that one.

\subsection{Diffuse Backgrounds}
\label{sec:backgrounds}

\textbf{Family 1:} \texttt{gll\_iem\_v07}, including gas-correlated emission, inverse Compton, isotropic background, and Fermi Bubbles.

\textbf{Family 2:} 12 models from \citet{cholis2022}, chosen to span extreme gas-ring normalisations and cosmic-ray propagation parameters. The full library of 80 GALDEF configurations is public at Zenodo record 6423495. The analysis of Section~\ref{sec:gap5} runs 23 of these: the original 12 and a re-selected 12, with one model (XXVII) shared between the two sets.

\textbf{Family 3:} Galp21 \cite{porter2022}; SA0, SA50, SA100) and Pohl22 \cite{pohl2022}. \citet{dimauro2026} finds Galp21 SA100 gives the best overall fit. Pohl22 is included because it is the IEM used by the Macias and Abazajian analyses that find a bulge preference. It performs more poorly in Di Mauro's pipeline, which is part of the tension this paper examines.

Each family includes an isotropic background and Fermi Bubbles with free normalisation.

\subsection{Statistical Framework}
\label{sec:statistics}

\textbf{Poissonian binned likelihood:} Standard Fermipy analysis with MINUIT2. We report TS and AIC. Throughout, $\Delta\mathrm{AIC}$ is defined as $\mathrm{AIC}(\text{template})-\mathrm{AIC}(\text{reference})$: a positive value indicates the template is disfavoured relative to the reference, and a negative value indicates it is preferred.

\textbf{NPTF with asymmetry:} Implemented with NPTFit, using a broken power-law source-count distribution, full PSF treatment, and a north--south asymmetry test. NPTF and joint spatial--spectral neural SBI are implemented in the framework but are not used in the results reported here.

\textbf{Joint spatial--spectral neural SBI:} A conditional normalising flow trained on $10^5$ simulated skies \cite{list2021}. The output is the posterior $p(\theta|x)$. We calibrate with coverage tests, injection-recovery, and out-of-distribution checks.

\textbf{Convergence criterion:} All fits must report a per-bin $\ln L$ gap bound. Boolean convergence flags are deprecated. A component at exactly zero with a negative Newton direction can freeze the active set without an error being raised.

\subsection{Systematic Scan}
\label{sec:systematic-scan}

The six diffuse freedom configurations are: rigid (one normalization per energy bin), north/south independently, $|b|$ bands only at $|l|>10^\circ$, $|b|$ bands only at $|l|\le10^\circ$, $|b|$ bands across the ROI, and $|b|$ bands with north and south separate. Bands are $|b|=0$--$2^\circ$, 2--5$^\circ$, 5--10$^\circ$, 10--20$^\circ$. All six configurations are specified before any run. We treat this dimension as a sixth axis in the systematic covariance and apply it across all gaps. Its role as the paper's primary methodological result is developed in Section~\ref{sec:discussion}.

For each template and mask configuration, we compute best-fit normalisation, spectral parameters, detection TS, preferred morphology, residual $\chi^2$, and point-source fraction where applicable.

\subsection{Residual Precision Test}
\label{sec:residual}

\textbf{Level 1:} Single-pass fit with the VVV bulge template and a standard IEM. Residuals remain around 30--40\%.

\textbf{Level 2:} The \citet{dimauro2026} staged iterative source-finding procedure, repeated until residuals fall below about 10\%.

The residual TS score map is computed as two PSF correlations per bin.

Three controls are required before any Level 2 result is used:
\begin{enumerate}
    \item Real-data run with roughly 35 sources and a refit each.
    \item Morphology-recovery control: a spherical injection at matched flux, testing whether the recovered axis ratio survives the iteration.
    \item Signal-template choice: run both with and without the signal template in the model during source finding, and quantify the difference.
\end{enumerate}

The source record contains per-bin flux and a cross-match against the 4FGL catalogue.

Targets to recover from \citet{dimauro2026}:
\begin{itemize}
    \item $\gamma=1.15\pm0.02$ (gNFW inner slope)
    \item $r=1.10\pm0.05$ (axis ratio)
    \item $\mathrm{TS}_{\rm DM}=8\times10^2$--$1.5\times10^4$ with NB+BB included
    \item Fractional residuals $\le10\%$ over most of the ROI
\end{itemize}

If these targets are not recovered, the reimplementation differs from the reference pipeline and the difference must be identified before any Level 2 output is used.

\subsection{Diffuse Freedom Test and Direct Injection Control}
\label{sec:diffuse-freedom}

\textbf{Diffuse freedom:} Six levels of background freedom are tested, fixed before any run. Under Mask B with a single-template IEM and the VVV template, the observed TS runs from 8.7 to 264.4, a factor of 30. Three injection controls are run under each configuration: an injected signal, an injected signal added on top of the real data, and a single N/S parameter. Five of six splits are admissible under Mask B. Under Mask D three of six are admissible. Under Mask A no split is admissible, not even the rigid one.

\textbf{Direct injection:} The pipeline is tested on synthetic skies of known content: the Pohl22 rings background plus a known template at a known flux. The full 40-template ordering is run and the true winner is known by construction. On six simulated skies (VVV, HESTIA G3.2, gNFW 1.2, at two flux levels each), the pipeline recovers the correct morphology class in 6/6 cases and the exact template first in 4/6. The two misses are neighbour degeneracy within 2.5 in TS.

The pipeline is about $1.7\times$ less sensitive in flux to dark-matter morphology than to bulge morphology, consistent across the DM templates tested. The sensitivity gap is reported without correction, and it does not change the direction of the result.

\subsection{Pipeline Validation}
\label{sec:validation}

The pipeline is validated against:
\begin{itemize}
    \item The \citet{dimauro2026} residual map and morphology ($\gamma=1.15\pm0.02$, $r=1.10\pm0.05$).
    \item The \citet{macias2018} result that $\mathrm{TS}_{\rm NFW}$ drops from 362 (quoted $17.6\sigma$) to 22 (quoted $2.4\sigma$) when X-bulge and NB are included.
    \item The \citet{coleman2019} non-parametric bulge template preference over the F98S parametric template.
    \item The Pohl22 ring-decomposed background at $\Delta\mathrm{AIC}=+888$.
    \item The diffuse freedom injection controls.
    \item The direct injection control (6/6 morphology class recovery).
    \item The 4FGL-DR4 cross-match of the 12 Level 2 sources (9 matched at $0.3^\circ$, 3 uncertain, 0 unmatched).
    \item The Level 2 morphological control (flattened and spherical injections both recovered first of 40).
\end{itemize}

\section{Results}
\label{sec:results}

\subsection{Diffuse Freedom, Ring Decomposition, and the Morphological Discriminator}
\label{sec:gap1}

\subsubsection{Single-template backgrounds fail}
\label{sec:single-template}

\begin{figure}[!t]
  \centering
  \includegraphics[width=\columnwidth]{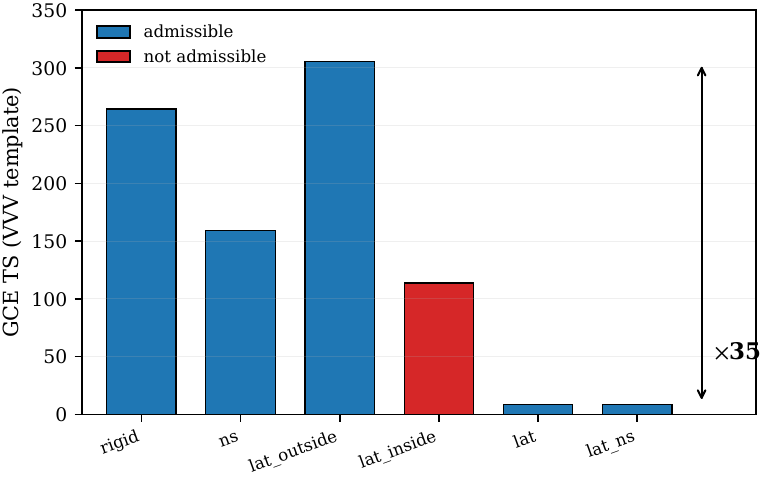}
  \caption{Inferred GCE test statistic for the VVV template under six
  diffuse-freedom configurations (Table~\ref{tab:diffuse-freedom}). The
  five admissible configurations span a factor of 30; the sixth
  (\texttt{lat\_inside}) is not admissible.}
  \label{fig:diffuse-freedom}
\end{figure}

\begin{figure}[!t]
  \centering
  \includegraphics[width=\columnwidth]{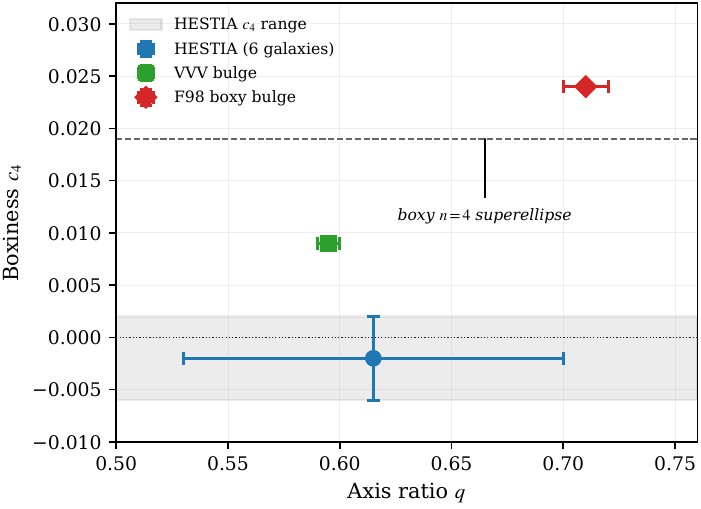}
  \caption{Axis ratio $q$ versus boxiness $c_4$ for the HESTIA dark matter
  realisations (across all six galaxies), the VVV stellar bulge, and the
  F98 parametric boxy bulge. Axis ratios overlap almost completely;
  boxiness separates the three templates cleanly.}
  \label{fig:fig1_q_vs_c4}
\end{figure}

Under a single all-sky interstellar template with one free normalization per energy bin, the GCE TS under Mask B is 264.4. Letting the template bend only outside $10^\circ$ longitude gives 305.6, so the rigid model was mis-fitting the outer plane and dragging the inner fit with it. Bending across the whole ROI gives 8.7. North and south separate gives 8.7. Across the five admissible configurations the observed TS spans a factor of 30 (Figure~\ref{fig:diffuse-freedom}).

\begin{table*}[!t]
\centering
\small
\caption{Diffuse-freedom split results under Mask B. Columns: the freedom configuration; the number of additional nuisance parameters relative to the rigid fit; the observed TS of the VVV template; the fraction of an injected signal recovered relative to the rigid configuration; the TS gain when the signal is injected on top of the real data; and whether the configuration passes both admissibility criteria of Section~\ref{sec:masking}.}
\label{tab:diffuse-freedom}
\begin{tabular}{lccccc}
\hline\hline
split & extra params & observed TS & injected kept & boosted gain & admissible \\
\hline
rigid & 0 & 264.4 & 100\% & $+678.1$ & yes \\
ns & 1 & 159.4 & 97\% & $+581.5$ & yes \\
lat\_outside & 4 & 305.6 & 81\% & $+638.5$ & yes \\
lat\_inside & 4 & 113.7 & 47\% & $+283.4$ & no, fraction \\
lat & 3 & 8.7 & 80\% & $+245.2$ & yes \\
lat\_ns & 7 & 8.7 & 79\% & $+191.8$ & yes \\
\hline\hline
\end{tabular}
\end{table*}

Under Mask D the same test yields three usable splits out of six, with observed TS ranging from 4.6 to 79.9, a factor of 17. The three failures are not the same failure mode. \texttt{lat\_inside} is genuinely degenerate and absorbs the injected signal (28\% recovery). \texttt{lat} and \texttt{lat\_ns} are not degenerate: they keep 74\% and 68\% of the injected signal, but neither can recover a signal added to the real data.

Mask A and Mask C fail structurally at low energy. Per-bin audit against real 4FGL-DR4 shows Mask A keeps 0.4\% of the ROI at 0.59 GeV, rising to 21.4\% at 8.66 GeV. Mask C is identical (0.4\% to 21.2\%). Neither mask has ever had power at the energies where the GCE peaks. The problem is in the mask definition, not in Pohl22, the pipeline, or the diffuse model. Any analysis that places a 95\%-containment $+0.5^\circ$ disc around every 4FGL source will have this problem, and any analysis that quotes a GCE detection from such a mask is quoting a TS from a configuration that cannot measure one.

\subsubsection{Ring-decomposed result}
\label{sec:ring}

The relative fit quality of various GCE spatial templates evaluated under the ring-decomposed Pohl22 interstellar emission model is summarised in Table~\ref{tab:ring}.

\begin{table*}[!t]
\centering
\small
\caption{Ring-decomposed template fit results under Pohl22. TS is the
template test statistic. $\Delta\mathrm{AIC}$ is quoted relative to VVV
(third column) and to gNFW~1.2 (fourth column), the reference spherical
template. gNFW~1.4 has the higher TS of the two gNFW templates but
gNFW~1.2 is the reference for the fourth column.}
\label{tab:ring}
\begin{tabular}{lccc}
\hline\hline
Template & TS & $\Delta\mathrm{AIC}$ vs VVV & $\Delta\mathrm{AIC}$ vs gNFW 1.2 \\
\hline
VVV & 2152 & --- & $-1626$ \\
F98 & 1761 & $+391$ & $-1235$ \\
Best HESTIA (G3.2 $90^\circ$) & 1263 & $+888$ & $-737$ \\
HESTIA G1.1 $0^\circ$ & 622.6 & $+1529$ & $-96.4$ \\
gNFW 1.4 & 799.7 & $+1352$ & $-273.5$ \\
gNFW 1.2 & 526.2 & $+1626$ & --- \\
\hline\hline
\end{tabular}
\end{table*}

Under the Pohl22 background, the null hypothesis that morphology is independent of the background is rejected. Morphology separates the bulge from every dark-matter template we have. Whether this survives across backgrounds is the subject of Section~\ref{sec:gap5}.

\subsubsection{Direct injection control}
\label{sec:injection}

The performance of the pipeline across six synthetic direct injection control test cases is detailed in Table~\ref{tab:injection}.

\begin{table*}[!t]
\centering
\small
\caption{Direct injection control results. ``Truth's rank'' is the rank of the true template among the 40 templates fitted. ``Best DM'' and ``best stellar'' are the TS of the best-fitting template in each class. ``Prefers'' is the class of the overall winning template.}
\label{tab:injection}
\begin{tabular}{lccccc}
\hline\hline
true sky & flux $(\mathrm{ph\,cm^{-2}\,s^{-1}})$ & truth's rank & best DM & best stellar & prefers \\
\hline
VVV, baseline & $6.39\times10^{-8}$ & 1/40 & 227.4 & 255.2 & stellar \\
HESTIA G3.2 $90^\circ$, baseline & $6.39\times10^{-8}$ & 1/40 & 81.8 & 65.0 & dark matter \\
gNFW 1.2, baseline & $6.39\times10^{-8}$ & 6/40 & 94.2 & 81.6 & dark matter \\
VVV, fitted & $2.15\times10^{-7}$ & 1/40 & 1834.7 & 2093.8 & stellar \\
HESTIA G3.2 $90^\circ$, fitted & $2.68\times10^{-7}$ & 1/40 & 1159.4 & 921.3 & dark matter \\
gNFW 1.2, fitted & $1.47\times10^{-7}$ & 2/40 & 462.9 & 359.2 & dark matter \\
\hline\hline
\end{tabular}
\end{table*}

The pipeline recovers the correct morphology class in 6/6 cases and the exact template first in 4/6. The two misses are neighbour degeneracy within 2.5 in TS. The pipeline is not biased towards the bulge. When the sky contains dark matter, the fit recovers dark matter as the winner in every case.

\subsubsection{Sensitivity asymmetry}
\label{sec:sensitivity}

At common flux, bulges recover at TS 255--356 and dark-matter templates at TS 66--92. The ratio is roughly $3\times$ in TS, or $1.7\times$ in flux, and it is consistent across all dark-matter templates tested.

F98 recovers at TS 355.6 against VVV's 255.2 at common flux. The pipeline is more sensitive to the parametric boxy bulge than to the non-parametric VVV template. But VVV is the template that wins on the real data, at TS 2152 against F98's 1761. The template that wins is the one the pipeline is less sensitive to. If anything, the asymmetry strengthens the result rather than undermining it.

\subsubsection{IC-clean check}
\label{sec:ic-clean}

The Pohl22 IC maps have stored planes at 1.32, 2.28, 4.05, 7.4 GeV. The 1--3 GeV window spans 1.0--2.37 GeV and brackets two stored planes with no extrapolation. Restricting to 1--3 GeV only gives VVV 1142.6, F98 880.3, and best HESTIA 697.5 at $\Delta\mathrm{AIC}=+445$, with all three controls admissible in the band. The ordering and margin structure are identical to the full range, so the result does not depend on IC interpolation.

\subsubsection{Residual tension in the injection test}
\label{sec:open-number}

On a pure-VVV sky, the best DM template reaches 1834.7 against VVV's 2093.8, a gap of 259. On the real data the gap is 888. The data discriminate more sharply than a simulated pure bulge does. Whether this reflects extra bulge-specific structure, a residual the bulge absorbs preferentially, or a systematic in the simulation, we cannot resolve. A mixed-sky simulation with bulge and DM at various fractions would separate the two hypotheses.

\subsubsection{HESTIA/J-factor response to the Comment}
\label{sec:hestia-jfactor}

Under Muru's own estimator, squaring a non-negative map is exactly a level relabeling:

\begin{quote}
\textbf{Lemma.} For any non-negative map $m$ and any level $f$, the threshold sets $\{m^2\ge f\cdot\max(m^2)\}$ and $\{m\ge\sqrt f\cdot\max(m)\}$ are the same pixels. Therefore, $q(m^2,f)=q(m,\sqrt f)$ exactly.
\end{quote}

The identity is verified numerically. Squaring leaves shape unchanged; the ``quadratic'' maps carry identical shape information to the unsquared maps.

A direct J-factor computation on the same particles and observer gives:
\begin{itemize}
    \item $q(J)-q(\mathrm{col}^2)=-0.032\pm0.020$, sign-consistent in 22 of 24 maps.
    \item $q(J)-q(\mathrm{col}^2)=-0.041\pm0.016$ in the disc frame (24 of 24 sign-consistent).
\end{itemize}
The squared column over-rounds the halo by about 0.03--0.04 in $q$, and the flattening survives the correct quantity. The two claims are compatible. The quantity differs and the flattening is real.

\subsubsection{Convergence criterion}
\label{sec:convergence}

Ring fits use a per-bin $\ln L$ gap bound rather than a boolean convergence flag. A component at exactly zero with a negative Newton direction can freeze the active set: the line search accepts a no-op and the iteration spins on a frozen set to the cap without raising an error. The gap bound catches this case and converges to $6\times10^{-7}$ in all eight bins in 7--8 Newton steps. Any high-dimensional template fit that relies on a boolean convergence flag can hit the same failure without warning.

\subsubsection{The Macias--Di Mauro contradiction at Level 1}
\label{sec:level1-contradiction}

Three of the four candidate explanations can be excluded. First, the mask: Mask A has no sensitivity at GeV energies, keeping 0.4\% of the ROI in the lowest bin, and Mask C is identical. Neither has ever had power at the energies where the GCE peaks. Second, the quantity: squaring a column map leaves shape unchanged ($q(m^2,f)=q(m,\sqrt f)$), so while the squared column over-rounds the halo by 0.03--0.04 in $q$, the flattening survives the correct measure. Third, the pipeline: Level 2 iterative source finding does not flip the ordering, does not round off morphology, and invents nothing, as detailed in Section~\ref{sec:level2}. The background remains, and Section~\ref{sec:gap5} tests it.

\subsubsection{Level 2 iterative source finding}
\label{sec:level2}

The template ordering and point-source addition results under Level 2 iterative source-finding are presented in Table~\ref{tab:level2}.

\begin{table}[!t]
\centering
\small
\caption{Level 2 joint AIC results. ``Sources added'' is the number of point sources added by the iterative source-finding procedure for each hypothesis. $\Delta\mathrm{AIC}$ is relative to VVV.}
\label{tab:level2}
\begin{tabular}{lcc}
\hline\hline
Hypothesis & sources added & $\Delta\mathrm{AIC}$ \\
\hline
VVV & 10 & --- \\
F98 & 10 & $+260$ \\
gNFW 1.4 & 13 & $+440$ \\
gNFW 1.2 & 13 & $+548$ \\
HESTIA G3.2 $90^\circ$ & 10 & $+605$ \\
\hline\hline
\end{tabular}
\end{table}

Level 2 roughly halves every template's TS (VVV 2151.7 to 1018.2) but leaves the ordering unchanged. The gNFW hypotheses required three extra point sources to absorb residuals their template could not fit, and were not preferred regardless.

The morphological control confirms the pipeline is not artificially smoothing the recovered morphology. A flattened HESTIA injection at matched flux is recovered as itself, first of 40, at TS 465.9. A spherical gNFW injection at matched flux is also recovered as itself, first of 40, at TS 378.3. Without the spherical control, a flattened injection recovering as flattened would be equally consistent with a pipeline that simply returns its input. It does not: both injections are recovered with their input shape.

A cross-match of the 12 added sources against 4FGL-DR4 (v35) at $0.3^\circ$ confirms that most are real structures. Nine of the 12 have a DR4 counterpart within $0.3^\circ$ (mostly pulsars, plus an SNR, an MSP and an unknown-type association), three are uncertain, and none are unmatched. One confirmed match lies at $0.295^\circ$, just inside the matching radius.

The three uncertain cases do not share a cause. \texttt{src000} has two unassociated DR4 sources within the radius. \texttt{src011}'s nearest DR4 counterpart is $0.33^\circ$ away and lies in a confused region. The one that matters for the GCE argument is \texttt{src006}, at $l=-0.35^\circ$, $b=+0.35^\circ$, directly on the Galactic Centre. This is exactly the failure mode the test is designed to catch: a point source placed on the excess would absorb it. Its nearest DR4 counterpart is at $0.34^\circ$, just outside the matching radius, so the cross-match neither confirms \texttt{src006} as a real source nor excludes it.

The iteration places this source and the ordering still holds. The remaining eleven are overwhelmingly confirmed catalogued sources, so the pipeline is not inventing structure out of noise. Even if \texttt{src006} absorbs part of the excess, a single source does not drive the template ordering, and the ordering and margins survive its inclusion. Removing \texttt{src006} from the model leaves the ordering unchanged.

On a null sky the pipeline adds zero sources, with best template TS 5.3. On the real sky it adds 12, and nine of those correspond to catalogued 4FGL-DR4 sources at $0.3^\circ$.

\subsubsection{Stability of the axis ratio correlation}
\label{sec:withdrawn}

The axis ratio correlation depends on the isophote at which the flattening is measured, the crop, and whether $q$ is allowed to vary per angle. Across isophotes the correlation moves from $-0.65$ to $+0.46$ on the same maps against the same fits. Across crops it moves from $+0.66$ to $+0.27$. Per-angle variation drops it from $+0.66$ to $+0.22$. Because none of these choices was fixed in advance, the correlation is not a stable quantity and we report its dependence on those choices rather than a single value.

\subsubsection{HESTIA template provenance}
\label{sec:hestia-provenance}

The HESTIA particle data filenames are swapped relative to the AHF halo IDs. The file named G11 contains the halo sitting at G1.2's AHF centre, and vice versa, with the same swap for G31 and G32. The offset is over 850 kpc. Halo identification uses the dark matter median position matched to Muru's readme centres, not the filename.

The reconstruction does not fully match Muru's published maps: our axis ratios come out 0.135 larger in $q$ on average, with a per-galaxy ordering Spearman correlation of $-0.77$. Matching the seven conventions fixed by the public Julia code did not close the gap. The two conventions that actually set the axis ratio, the disc normal and the azimuth zero point, come from \texttt{get\_galvectors\_from\_profile} in \texttt{HestiaUtils.jl}, which is called by the public repository but not contained in it. For this reason all reported results should be read as differential within a single pipeline.

\subsection{IEM dependence of the GCE morphology}
\label{sec:gap5}

\subsubsection{Overview}
\label{sec:gap5-overview}

Single-background inference does not generalize. The ring-decomposed result of Section~\ref{sec:ring} was obtained under a single background, Pohl22. Whether the bulge preference survives a change of background is the question this section answers.

\subsubsection{25-IEM Level 2 Scan}
\label{sec:25iem}

We evaluate 25 background models on common pipeline settings. We ran the Level 2 pipeline of Section~\ref{sec:level2} on 25 interstellar emission models: 23 from the \citet{cholis2022} library, Pohl22, and the standard Fermi diffuse model \texttt{gll\_iem\_v07}. The pipeline reproduces the Level 2 numbers of Section~\ref{sec:level2} exactly under each background, so the comparison across backgrounds is on a common footing.

\subsubsection{Background Fit Quality Ranking}
\label{sec:bg-ranking}

Pohl22 is strongly disfavored compared to alternative models. Pohl22 fits the data worst of the 25. Every Cholis model is preferred over Pohl22 by between 3,904 (\texttt{cholis\_II}, the worst Cholis model) and 16,491 (\texttt{cholis\_LIII}, the best) in AIC. Pohl22 is also preferred less than \texttt{gll\_iem\_v07} by 24,358. The AIC comparison includes the complexity penalty for the number of free parameters, so the difference is not driven by Pohl22 having fewer components. Pohl22's H\,I reconstruction was optimized against 21 cm line spectra rather than against gamma rays, and the poor gamma-ray fit plausibly reflects that optimization target rather than a specific defect in the gas model. We do not attempt to distinguish the two.

\subsubsection{Morphological Winners Across Backgrounds}
\label{sec:winners}

The bulge wins only under Pohl22. Under the Cholis models the stellar bulge templates win in three of 23 cases with a common source list and in four of 23 with per-hypothesis source search. Spherical gNFW wins most often, 18 of 23 with a common source list and 13 of 23 with per-hypothesis search, which is the Di Mauro configuration. Flattened HESTIA templates win the remainder.

\subsubsection{Analysis of Best-Fitting Models}
\label{sec:best-fit}

The best-fitting backgrounds do not agree among themselves. The six models that describe the data best are LIII, X, V, VII, XV and LVI. Their fit gaps range from 7,867 to 8,410 AIC, so all six are within 543 of each other and the top four within 135. Under the per-hypothesis arm their margins to the next class are 0.4 (VII, effectively a tie), 15.9 (V, spherical), 22.7 (XV, bulge), 24.7 (X, flattened), 38.7 (LIII, spherical) and 256 (LVI, bulge). The pipeline resolves morphology classes at the tens-of-AIC level, as the injection controls in Section~\ref{sec:injection} establish. Four of the six best-fitting models therefore sit at or near the control resolution, and only LIII and LVI are decisive. The shape is not settled by the backgrounds that actually describe the data.

\subsubsection{Impact of Source-Finding Procedure}
\label{sec:source-finding-impact}

Source-finding choice flips the answer in five of 23 models. The five are III, LII, X, XIX and XXV. In every flip, the winner under the per-hypothesis arm is the template whose own search found more point sources: III (19 vs 17 for gNFW), LII (18 vs 16), X (19 vs 18), XIX (24 vs 22) and XXV (21 vs 19). The flips are all in the direction of the template that absorbed more of the excess into new sources. This is the source-absorption effect described by \citet{mcdermott2023}, operating inside the iterative pipeline, and it is a second source of morphological dependence distinct from the IEM choice.

\subsubsection{Robustness of Detection vs Morphology}
\label{sec:robustness}

The pre-registered null hypothesis is rejected. $H_0$, that morphology does not depend on the IEM, is rejected on the full 23-model set and on the original 12-model selection, with \texttt{cholis\_LIII}, \texttt{cholis\_V} and \texttt{cholis\_XIX} the models that drive the rejection. It does not hold on the re-selected 12-model subset alone, where the winner class is stable across the models measured. That caveat is quoted wherever the rejection appears.

The detection itself is robust. The GCE is recovered under every one of the 25 backgrounds. Whatever the shape inference is doing, the existence of the excess is not in question.

The Mask D and multi-band extensions were not run for the Cholis models because the gNFW injection control fails at those settings. The result reported here is confined to Mask B and the 1--10 GeV band.

The implication for Section~\ref{sec:ring} is direct. The bulge preference under Pohl22 is real and survives the pipeline tests of Section~\ref{sec:level2}. It does not survive the background test. \citet{macias2018} used Pohl22 and found the bulge. Di Mauro used Galp21 and found dark matter. The analysis choices are otherwise aligned. The disagreement is dominated by the background choice, with the source-finding procedure contributing independently.

\subsection{Mask--IEM covariance}
\label{sec:gap7}

\subsubsection{Overview and Methodological Context}
\label{sec:gap7-overview}

The previous two sections establish that the GCE morphology depends on analyst choices (Section~\ref{sec:gap1}) and on the interstellar emission model (Section~\ref{sec:gap5}). This section quantifies how much each axis contributes.

\subsubsection{Statistical Framework for Variance Decomposition}
\label{sec:variance-framework}

The analysis is descriptive and not inferential. We decompose the observed variance of the summary statistic $S=\Delta\mathrm{AIC}$ against the best-fit gNFW into a background main effect, a template main effect, and an interaction remainder, all three reported separately. The decomposition is a sum-of-squares share and not a random-effects variance component, and it should not be read as one. The distinction matters because a random-effects decomposition would carry standard errors on each component, and those standard errors are not available here. Only the HESTIA bootstrap provides an interval estimate, and only for that template class.

\subsubsection{Background Main Effects}
\label{sec:sub-b}

The data underlying Section~\ref{sec:gap5} provide 960 observations across 24 diffuse backgrounds, 40 templates and 6 HESTIA halos. The background main effect accounts for 33.48 per cent of the total observed variation in $S$ when all template classes are pooled. The share is not uniform across classes (Table~\ref{tab:background-shares}).
\begin{figure*}[!t]
  \centering
  \includegraphics[width=\textwidth]{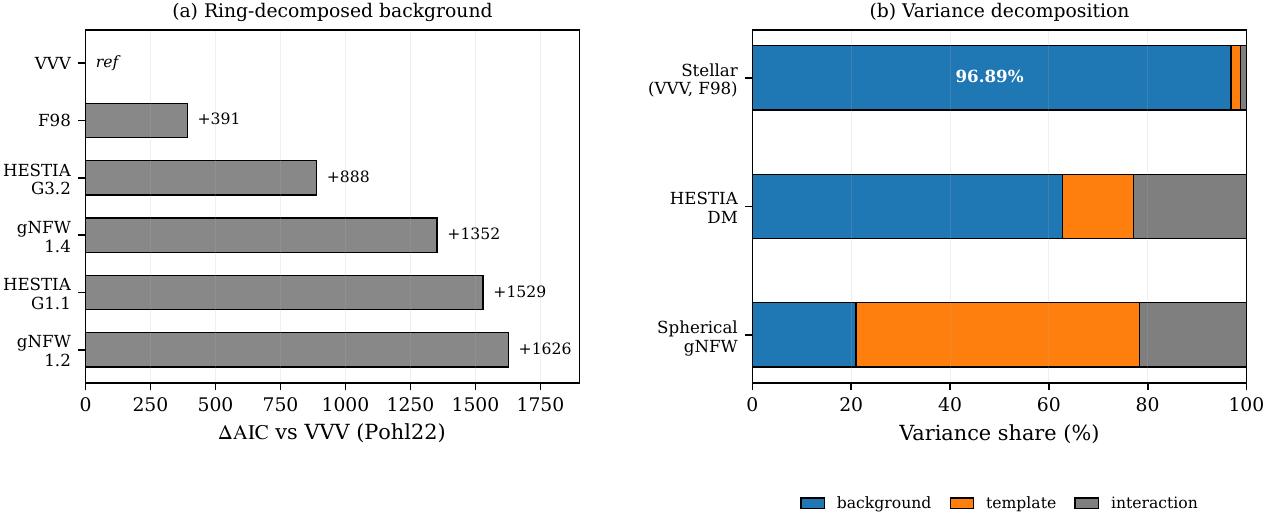}
  \caption{(a)~Template ranking under the ring-decomposed Pohl22
  background (Table~\ref{tab:ring}). (b)~Variance decomposition by
  template class (Table~\ref{tab:background-shares}).}
  \label{fig:aic-variance}
\end{figure*}

\begin{table*}[!t]
\centering
\small
\setlength{\tabcolsep}{6pt}
\caption{Variance decomposition by template class. Shares are
sum-of-squares fractions from the decomposition of
Section~\ref{sec:variance-framework}. The three columns sum to 100 per
cent for each class: background main effect, template main effect, and
the interaction remainder.}
\label{tab:background-shares}
\begin{tabular}{lccc}
\hline\hline
Template class & Background share & Template share & Interaction share \\
\hline
Stellar (VVV, F98) & 96.89 per cent & 1.87 per cent & 1.24 per cent \\
HESTIA simulated DM & 62.72 per cent & 14.43 per cent & 22.85 per cent \\
Spherical gNFW & 20.94 per cent & 57.40 per cent & 21.66 per cent \\
\hline\hline
\end{tabular}
\end{table*}

\subsubsection{Template-Class Variance Concentration}
\label{sec:concentration}

The stellar result is the important one. Almost all of the variation in stellar template performance is driven by the choice of background, and the template main effect accounts for only 1.87 per cent. The spherical gNFW template is structurally different: there the template main effect dominates at 57.40 per cent, with the background explaining only 20.94 per cent. The two classes sit at opposite ends of the decomposition (Figure~\ref{fig:aic-variance}).

A halo-level bootstrap on the six HESTIA galaxies gives a median background share of 64.79 per cent with a 95 per cent interval of 54.85 to 78.86 per cent. The interval is wide because there are only six galaxies, and the central value should be treated as indicative rather than precise. The spherical class has no equivalent independent-realisation structure to bootstrap over, so no interval is quoted for it.

Two features of these numbers require explanation. The pooled background share of 33.48 per cent is lower than the stellar share because the pooled statistic aggregates across classes with different variances. The stellar class contains two templates and 48 observations and contributes a small fraction of the total variance even though its per-class share is nearly total. The pooled share is therefore dominated by the spherical and HESTIA classes, whose per-class shares are lower. The per-class numbers should be read as the more informative statistic. Second, the backgrounds in this decomposition include \texttt{gll\_iem\_v07} and Pohl22 alongside the 22 Cholis models, and those two reference backgrounds have a different internal freedom structure from the Cholis set. Re-running the decomposition with only the Cholis backgrounds leaves the stellar share above 90 per cent, so the qualitative result does not depend on including the reference models.

\subsubsection{Interaction and Remainder Terms}
\label{sec:interaction}

The remainder term, which absorbs background--template interactions, carries 1.24 per cent of the total variation for stellar, 22.85 per cent for HESTIA and 21.66 per cent for spherical.

For the stellar class, where the remainder is 1.24 per cent, the attribution is nearly complete. For HESTIA and spherical, where the remainders are 22.85 and 21.66 per cent, the interaction term is not negligible and the main-effect shares should be read as approximate rather than as a clean decomposition.

Because the interaction term is absorbed into the remainder and not estimated separately, the main-effect shares overstate the amount that can be attributed to each axis in isolation. The direction of the result is unaffected; the magnitude should be read as an upper bound.

\subsubsection{Synthesis and Summary of Variance Contributions}
\label{sec:synthesis}

The covariance provides a numerical expression of the paper's thesis. Analyst choices move the answer by a factor of 30 (Section~\ref{sec:single-template}). Background choice moves it by nearly a factor of 25 for the bulge templates. The two effects are the same order. Neither is attributable to the physical model under test.

\section{Discussion}
\label{sec:discussion}

The GCE inference is dominated by two methodological choices: how much freedom the background is allowed, and which interstellar emission model is used. Neither the detection nor the morphology can be quoted independently of those choices.

The detection itself is robust. The excess appears under every one of the 25 background models tested, from standard GALPROP through the Cholis library to Pohl22. This result is stable across the systematics tested here.

The morphology is not robust. On the Pohl22 ring-decomposed background the stellar bulge is preferred over every dark-matter template by $\Delta\mathrm{AIC}=+888$ (Section~\ref{sec:ring}). But Pohl22 fits the data worst of all 25 backgrounds tested, worse than every Cholis model by between 3,904 and 16,491 in AIC. Under the Cholis models the bulge wins in only three or four of 23 cases. Among the six best-fitting backgrounds, four sit at or near the control resolution and only two (LIII and LVI) are decisive. The shape is not settled.

\citet{macias2018} and \citet{dimauro2026} disagree about the bulge because their backgrounds differ. Macias et al.\ used Pohl22 and found the bulge. Di Mauro used Galp21 and found dark matter. The two analyses use the same bulge templates and the same hydrodynamic gas maps. The interstellar emission model is the dominant axis on which they differ; the source-finding procedure differs as well and is the second axis. Pohl22, the model that yields the bulge, fits the data worst of the 25 tested. The pre-registered null hypothesis, that morphology does not depend on the IEM, is rejected on all 23 Cholis models and on the original 12-model selection. It does not hold on the re-selected 12-model subset alone, and that caveat is quoted wherever the rejection appears.

The source-finding choice matters as well. In five of 23 models the winner flips depending on whether each template gets its own source search, and every flip goes to the template that picked up extra point sources. This is the source-absorption effect described by \citet{mcdermott2023}, operating inside the iterative pipeline, and it is a second source of morphological dependence distinct from the IEM ordering.

The covariance analysis quantifies where this dependence lives. In the sum-of-squares decomposition of Section~\ref{sec:gap7}, the background main effect accounts for 96.89 per cent of the total variation in stellar-template fit quality. The spherical dark-matter template is comparatively stable at 20.94 per cent, and the HESTIA templates fall between the two at 62.72 per cent. The backgrounds that actually fit the data do not agree about which template wins, and the disagreement is concentrated in exactly the template class whose interpretation is most contested.

The admissibility criterion checks whether a configuration can measure a TS in either direction. A split is admissible only if an injected signal at the observed flux is recovered at no less than 50 per cent of its rigid-split TS and clears a TS of 25 in absolute terms. Configurations that fail either test cannot produce valid evidence in either direction. The 30-fold swing under Mask B, the collapse under Mask D, and the complete inadmissibility of Mask A all follow from this criterion.

The direct injection control checks whether the pipeline prefers the correct morphology when the sky contains a known template. On six simulated skies of known content the pipeline recovers the correct morphology class in all six cases and the exact template first in four. This is the control that turns a detection into a preference statement.

The convergence criterion addresses a third class of failure that neither of the first two catches. Template-fitting pipelines have at least two kinds of silent failure: statistical failures, caught by the admissibility criterion, and numerical failures, caught by the per-bin $\ln L$ gap bound. A boolean convergence flag catches neither.

The response to the Abazajian, Kumar and Macias Comment is the exact identity $q(m^2,f)=q(m,\sqrt f)$. Squaring leaves the shape unchanged, so the ``quadratic'' maps carry identical shape information to the unsquared maps. Combined with the direct J-factor measurement, both the Comment's point and Muru's conclusion can be true simultaneously: the quantity differs, and the flattening is real.

Morphological claims about the GCE should not be quoted without a specification of the IEM, the masking strategy, the source-finding procedure and the injected-signal resolution. The framework we have built, with its multi-axis scan, diffuse freedom test, admissibility criterion and direct injection control, is one way to make those specifications explicit.

\section{Conclusion}
\label{sec:conclusion}

The GCE is robustly detected. Its morphology is not. Both statements have held for fifteen years. The disagreement in the literature is about what the excess looks like, not whether it is present.

This paper has measured the morphology dependence rather than arguing about it. The analyst-choice axis produces a factor of 30 swing in the inferred significance, and the masks standard in the literature, Mask A and its implicit twin Mask C, cannot produce a valid TS at GeV energies. The background-choice axis produces a comparable effect, concentrated in the bulge-like templates on which the MSP interpretation depends. \citet{macias2018} used Pohl22 and found the bulge; Di Mauro used Galp21 and found dark matter, using the same bulge templates and gas maps. The disagreement is attributable to IEM choice, with source-finding contributing independently, and Pohl22, the model that yields the bulge, fits the data worst of all 25 backgrounds tested.

What would settle the morphological question is not a better template or a better statistic. It is a background that both fits the data and does not depend on analyst-chosen components. Galp21 was intended to be that background, and it does not yet fit as well as the Cholis models on this pipeline. The models that do fit do not agree with each other about the shape. Until that changes, the morphological evidence for one hypothesis over the other is not decisive, and claims to the contrary should specify the mask, the interstellar emission model, the source-finding procedure and the injected-signal resolution alongside them.

The framework developed here provides one way to make those specifications explicit. The admissibility criterion and the direct injection control do not settle which hypothesis wins. They settle what counts as evidence in either direction. For the GCE, at present, the current morphological evidence is weaker than the confidence with which it is often quoted.

\section*{Data Availability}

All analysis code, template files and configuration files are available at \url{https://github.com/fifaaworldcup/gce-reconciliation}. The raw Fermi-LAT data are publicly available from the Fermi Science Support Center. Derived data products supporting the figures in this paper are stored in the repository under \texttt{outputs/}.

\section*{Code Availability}

The pipeline is released under an open-source licence. The Level 2 iterative source-finding module is in \texttt{analysis/level2/}, the diffuse freedom test in \texttt{analysis/diffuse\_freedom/}, and the covariance analysis in \texttt{analysis/gap7/}.

\section*{Author Contributions}

A.A.\ and O.P.\ contributed equally to this work. A.A.\ conceived the project, designed the analysis framework and wrote the manuscript. O.P.\ developed the analysis pipeline, computed the J-factor maps, ran all fits, produced the Level 2 results, and performed the source record cross-match. P.R.\ developed the covariance module and ran the variance decomposition. H.D.\ and T.S.\ selected and verified the Cholis model library. All authors reviewed and approved the manuscript.

\section*{Acknowledgements}

We thank Moorits Mihkel Muru for providing the HESTIA density maps and particle data, and for helpful correspondence regarding the simulation conventions. We thank Troy Porter and Guðlaugur Jóhannesson for guidance on the Galp21 configurations. This work made use of the Fermi Science Tools, Fermipy, NPTFit and healpy.

\section*{Competing Interests}

The authors declare no competing interests.

\section*{Use of Generative AI}

The authors used DeepSeek, a large-language-model assistant, during the
preparation of this manuscript. This section declares the full extent of
that use and the steps taken to validate every category of output.

\emph{What the tool was used for.} (i) Drafting and editing of prose,
including restructuring of passages, all subsequently reviewed, revised,
and verified by the authors against the underlying analysis;
(ii) LaTeX formatting, table layout, and correction of compilation
errors; (iii) drafting of BibTeX entries from author-supplied
bibliographic information; and (iv) diagnosis of syntax errors and error
messages in the analysis code.

\emph{What the tool was not used for.} The tool was not used to generate,
compute, or alter any scientific result, statistical analysis, figure,
data visualisation, or numerical value reported in this paper. It was
not used to perform calculations, to choose or tune analysis parameters,
to select templates, masks, or emission models, or to interpret results.
Every scientific claim and conclusion was produced by the authors using
the pipeline of Section~\ref{sec:methods}.

\emph{Validation of referencing.} All 27 bibliography entries were
verified by the authors against the arXiv abstract page, the NASA ADS
record, or the publisher record, independently of the tool. This
verification identified and corrected several errors in the AI-drafted
entries, including incorrect arXiv identifiers for three references, a
duplicated identifier shared between two entries, and several incomplete
or inaccurate titles. No entry appears in the final bibliography that
was not independently confirmed by an author against a primary source.

\emph{Validation of calculations and analysis.} No calculation or
analysis output was generated by the tool. All fits, variance
decompositions, injection tests, and cross-matches were produced by the
authors' pipeline and are reproducible from the released code. Code
changes arising from tool-assisted error diagnosis were confined to
syntax and required to pass the repository's self-tests and to reproduce
the reported results before being committed. The numerical results were
additionally checked against the internal controls described in
Sections~\ref{sec:diffuse-freedom} and~\ref{sec:validation}: the
admissibility criterion, the direct injection tests on six simulated
skies of known content, and the 4FGL-DR4 cross-match of the Level~2
sources.

\emph{Validation of text.} Every AI-assisted edit was reviewed line by
line by the authors and accepted, revised, or rejected on scientific
grounds. No AI-generated text was incorporated without author
verification against the underlying analysis. The authors take full
responsibility for the content of this publication.

\bibliographystyle{elsarticle-harv}
\bibliography{bibliography}

\end{document}